\documentclass{epsconf}
\usepackage{graphicx}
\usepackage{wrapfig}
\usepackage{amsmath}
\usepackage[english]{babel}
\usepackage{color}

\newcommand{\omp}{\omega_p}
\newcommand{\p}{\partial}
\newcommand{\reff}[1]{(\ref{#1})}
\newcommand{\eref}[1]{Eq.\reff{#1}}
\newcommand{\erefs}[1]{Eqs.\reff{#1}}
\newcommand{\figref}[1]{Fig.\ref{#1}}

\title{Hierarchical approach for energetic particle transport \\ in 1-dimensional uniform plasmas}
\author{\underline{N. Carlevaro}$^{1,2}$, F. Cianfrani$^{3}$, {G. Montani}$^{1,4}$, F. Zonca$^{1,5}$}
\institute{$^{1}$ ENEA, FNS Department, C.R. Frascati,
Via E. Fermi 45, 00044 Frascati (Roma), Italy\\
$^2$ CREATE Consortium, Via Claudio 21 (80125) Napoli, Italy\\
$^3$ PIIM UMR7345, CNRS, AMU, Jardin du Pharo, 58 Bd C. Livon, 13007
10 Marseille, France\\
$^4$ Physics Department, ``Sapienza'' University of Rome, P.le Aldo Moro 5, 00185 Roma, Italy\\
$^5$ IFTS and Department of Physics, Zhejiang University, Hangzhou 310027, China
}

\begin{document}
\maketitle

\vspace{-0.8cm}
\paragraph{Abstract}
The importance of the beam-plasma system (BPS) in fusion physics relies on its capability in reproducing relevant features of energetic particles interacting with the Alfv\'enic spectrum \cite{cz,bs}. We analyze here a multi-level hierarchy of the Vlasov-Poisson (VP) induced transport in order to characterize the underlying physical processes.

\vspace{-5mm}
\paragraph{Hamiltonian description of the beam-plasma interaction}
The BPS faces the resonant dynamics of a fast particle beam injected into a 1D plasma, which is treated as a cold linear dielectric medium supporting electrostatic turbulence. We adopt the Hamiltonian formulation\footnote{Notation: The 1D cold plasma is taken as a periodic slab of length $L$. Beam particle positions and velocities are $x_i$ and $v_i$, $N$ is the total particle number. The  electrostatic potential $\varphi(x,t)$ is expressed in terms of the Fourier components $\varphi_k(t)$ ($k$ is the wave-number). Introducing the beam to plasma density ratio $\eta=n_B/n_p\ll1$, we use the dimensionless variables: $\bar{x}_i=x_i(2\pi/L)$, $\tau=t\omp$, $u_i=\bar{x}_i'=v_i(2\pi/L)/\omp$, $\ell=k(2\pi/L)^{-1}$ (integers), $\phi_\ell=(2\pi/L)^2 e\varphi_k/m\omp^2$ and $\bar{\phi}_\ell=\phi_\ell e^{-i\tau}$. The prime denotes $\tau$ derivative.} of the problem \cite{jpp} where the broad energetic particle beam self-consistently evolves in the presence of $M$ linearly unstable modes, each one almost at the plasma frequency, i.e. $\omega\simeq\omp$:
\begin{align}\label{mainsys1}
\bar{x}_i'=u_i \;,\qquad
u_i'=\sum_\ell\big(i\,\ell\;\bar{\phi}_\ell\;e^{i\ell\bar{x}_{i}}+c.c.\big)\;,\qquad
\bar{\phi}_\ell'=-i\bar{\phi}_\ell+\frac{i\eta}{2\ell^2 N}\sum_\ell e^{-i\ell\bar{x}_{i}}\;.
\end{align}
The resonance conditions write $\ell u_{r}=\omega/\omp\simeq1$ ($u_r$ being the resonant velocities) and the warm beam is initialized with an assigned distribution function (DF) $\bar{F}_B(u)$, with $S=\int du \bar{F}_B(u)$.

\vspace{-5mm}
\paragraph{Vlasov-Poisson system}
The BPS can be treated kinetically via the VP coupled system expressed using the Fourier components of the electric field ($E_{k}(t)$) and of the beam DF ($f_k(t,v)$):
\begin{align}
\p_t f_k = -ikv\,f_k + \frac{e}{m}\sum_{k'} E_{k'}\p_v f_{k-k'}\;,\qquad
\p_t E_k = -i\omega _pE_k + \frac{2\pi e\omega _p}{k}\int_{-\infty}^{\infty}\!\!\!\!dv f_k\;.
\label{a1}
\end{align}
Due to the initial spatial homogeneity of the system, $f_0\equiv f_B(t,v)/L$ is the only $k$ having non-zero initial conditions and it is governed by the following dimensionless transport equation:
\begin{align}\label{dbnksdkjb}
\p_\tau\bar{f}_{B}(\tau,u)=\p_u\,\Gamma_{bps}(\tau,u)=\p_u\Big[4\pi\sum_\ell\Big[
\ell\bar{\phi}_\ell^b\,\bar{f}_{\ell}^a-\ell\bar{\phi}_\ell^a\,\bar{f}_{\ell}^b
\Big]\Big]\;,
\end{align}
where we used $u=1/\ell$, $E_k=-ik\bar{\varphi}_k=k\bar{\varphi}_k^b-ik\bar{\varphi}_k^a$ (here, $\bar{\varphi}_k=\bar{\phi}_\ell (m\omp^2/(2\pi/L)^2 e)$) and $\bar{f}(\tau,u,\bar{x})$ is taken from the histogram of the phase-space $N$-body simulations. \eref{dbnksdkjb} corresponds to the zeroth level of the hierarchy scheme we are analyzing: it allows to define the proper form of the fluxes $\Gamma_{bps}$, evaluated by sampling $\bar{\phi}_\ell(\tau)$ and $\bar{f}_\ell(\tau,u)$ from simulations of \eref{mainsys1}, to be compared with the other approximation levels defined in what follows. The evolution of the DF matches exactly the profiles obtained from the fully self-consistent scheme.

\vspace{-5mm}
\paragraph{Diagonal reduced Vlasov-Poisson system}
The single function $f_k$ is assumed to receive mainly contribution from the correspondent harmonics ($k'=k$ in \eref{a1}). 
It thus satisfies:
\begin{align}\nonumber
\p_t f_k = -ikv\,f_k + \frac{e}{m}E_k\p_v f_0\;,\qquad\Rightarrow\qquad
f_k(t,v)=\frac{e}{m}\int_0^t\!\!\! dt' E_k(t') e^{ikv(t'-t)}\p_v f_0(t',v)\;,
\end{align}
obtaining a diagonal reduced transport equation for $f_0$ (here and in the following, we use $k>0)$:
\begin{align}\label{a55}
\p_t f_0(t,v)-\frac{e^{2}}{m^{2}}\sum_{k}\Big[E_k\;\p_v\Big(\int_{0}^{t}\!\!\!dt' E_k^*(t')e^{ikv(t'-t)}\p_v f_0(t',v)\Big)+c.c.\Big]=0\;.
\end{align}
The electric field can be set as
$E_k(t') = E_k(t)\,\exp\big[-i\int_t^{t'}\!\!\!dt''\omega_k(t'')\big]$ getting, form \eref{a55} and \eref{a1} (right), a Dyson-like system for the evolution of $f_0$ and of the spectrum:
{\small
\begin{align} 
\p_t f_0(t,v)=\frac{e^2}{m^2}\sum_{k}|E_k(t)|^2
\p_v\Big[\int_0^t\!\!\! dt' \exp\Big(ikv(t'-t)-i\int_t^{t'}\!\!\! dt''\,\omega_k(t'')\Big) \p_v f_0(t',v)+c.c.\Big]\;,\label{sq0}\\
\p_t|E_k|^2=\frac{2\pi e^2\omega_p}{mk}|E_k(t)|^2
\Big[\int_{-\infty}^{\infty}\!\!\!\!dv\int_0^t\!\!\! dt' \exp \Big(ikv(t'-t)-i\int_t^{t'}\!\!\! dt''\omega_k(t'')\Big) \p_v f_0(t',v)+c.c.\Big]\;.
\label{sq}
\end{align}}

\vspace{-9mm}
\paragraph{A - External spectrum sampling (ES)}
\eref{a55} can be integrated for a given spectral evolution extracted from simulations. Using $G_\ell=\int_0^\tau d\tau' e^{i\ell u \tau'}\bar{\phi}_\ell\p_u\bar{f}_B$ and $\bar{G}_\ell=e^{-i\ell u \tau}G_\ell$, we get
\begin{align}\label{deqexf}
\p_\tau \bar{f}_B(\tau,u)=\p_u\,\Gamma_{es}(\tau,u)=\p_u \sum_\ell\ell^2 (\bar{\phi}_\ell \bar{G}_\ell^*+\bar{\phi}_\ell^* \bar{G}_\ell)\;,\quad
\p_\tau \bar{G}_\ell(\tau,u)= -i\ell u \bar{G}_\ell+\bar{\phi}_\ell\p_u \bar{f}_B\;.
\end{align}
A $4th$ order Runge-Kutta algorithm evolves the system with $\bar{f}_B(0,u)=\bar{F}_B(u)$ and $\bar{G}_\ell(0,u)=0$. \eref{deqexf} represents the first level of the hierarchy scheme: the fluxes $\Gamma_{es}$ can be now evaluated by sampling only $\bar{\phi}_\ell(\tau)$ from \eref{mainsys1} obtaining an approximated evolution of $\bar{f}_B(\tau,u)$.

\vspace{-5mm}
\paragraph{B - Quasi-linear model (QL)} The QL model, due to the specific underlying assumptions, corresponds to the second hierarchy level of the approximation scheme. The model results in a system of self-consistent equations for the DF evolution (no sampling from simulations). \erefs{sq0}-\reff{sq} can be reduced using the following assumptions: quasi-stationarity of $\omega_k$ and $f_0$; marginal stability for $\mathrm{Im}(\omega_k)\ll\omp$; broad and dense spectrum, i.e. continuous $k$-space $k=\omp/v$ (we can use $E(t,k)\to E(t,v)$). Introducing the spectral function $\mathcal{I}(\tau,u)=|\bar{\phi}|^2$ (with $\mathcal{I}_0=\mathcal{I}(0,u)$) and $\bar{\mathcal{N}}\!\!=\!\!M/(\ell_{max}\!-\!\ell_{min})$, $H(\tau,u)=(\pi\eta u^2/S)\int_0^\tau \p_u \bar{f}_Bd\tau'$, QL equations write
\begin{align}\label{QL_u}
&\p_\tau \bar{f}_B(\tau,u)=\p_u\Gamma_{ql}(\tau,u)=\p_u \Big[
\pi \bar{\mathcal{N}}\p_u \bar{f}_B\,\mathcal{I}_0\;\exp[H]/u^{3}\Big]\;,\quad
\p_{\tau}H(\tau,u)=\pi\eta u^2\,\p_u \bar{f}_{B}/S\;.
\end{align}
Initial conditions are $\bar{F}_B(u)$, $H(0,u)=0$ and the spectral evolution reads $\mathcal{I}_{QL}(\tau,u)=\mathcal{I}_0\;\exp[H]$.


\vspace{-5mm}
\paragraph{C - Extension of QL model}
The QL system 
can be re-derived \cite{ql} using an expansion for the DF at short times. 
This formally extend the validity of the QL model to the temporal mesoscales before saturation getting the following spectral correction:
\begin{align}\label{snknlkn}
\mathcal{I}=\mathcal{I}_{QL}\big(|\p_u \bar{F}_B|/|\p_u \bar{f}_B|\big)^{\alpha}\;,\qquad \alpha=4(\sqrt{2}-1)/\pi\simeq0.51\;.
\end{align}
We recognize this model as a level 2.0 of the hierarchy scheme. Here, the spectral correction (\eref{snknlkn}) is evaluated by using $\bar{f}_B$ sampled from simulations of the $N$-body scheme \eref{mainsys1}.

\begin{figure}[ht!]
\centering
\includegraphics[width=.33\textwidth,clip]{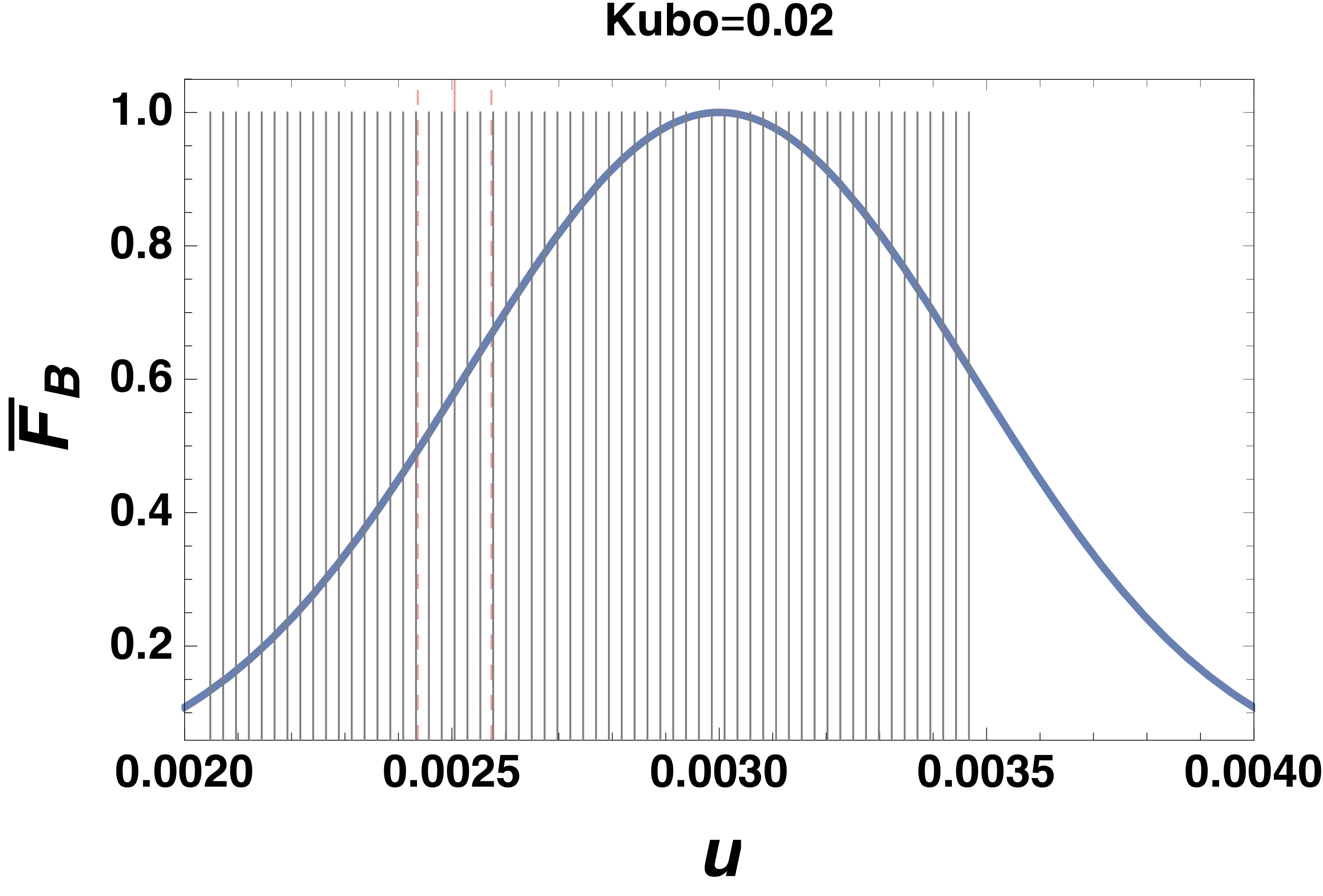}
\includegraphics[width=.32\textwidth,clip]{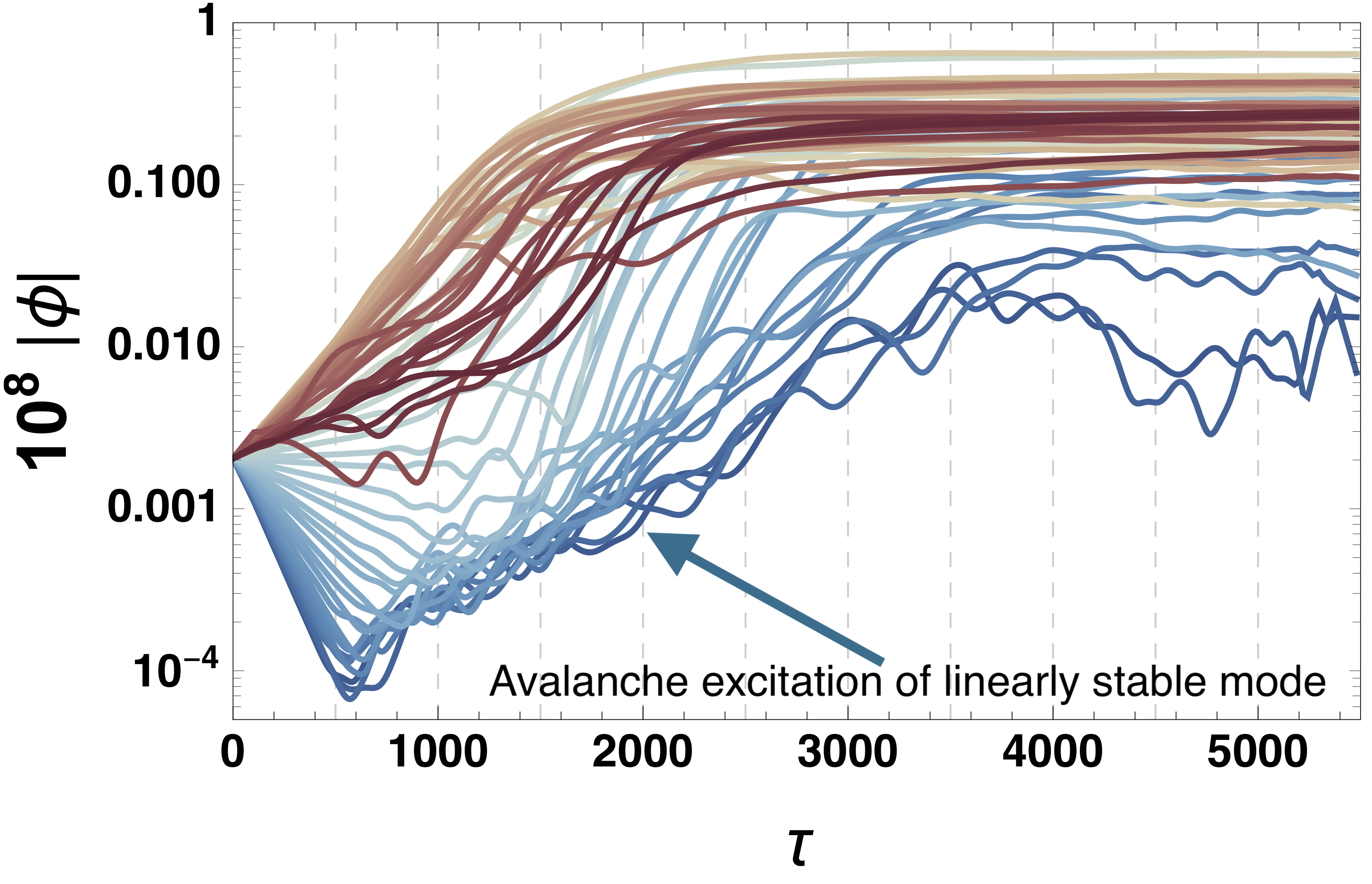}
\vspace{-7mm}
\caption{\it \footnotesize Left: initial profile and resonances. Right: mode evolution from \eref{mainsys1} (linear stable modes in blue). 
\label{figsetup}}
\end{figure}

\begin{figure}[ht!]
\centering
\includegraphics[width=.32\textwidth,clip]{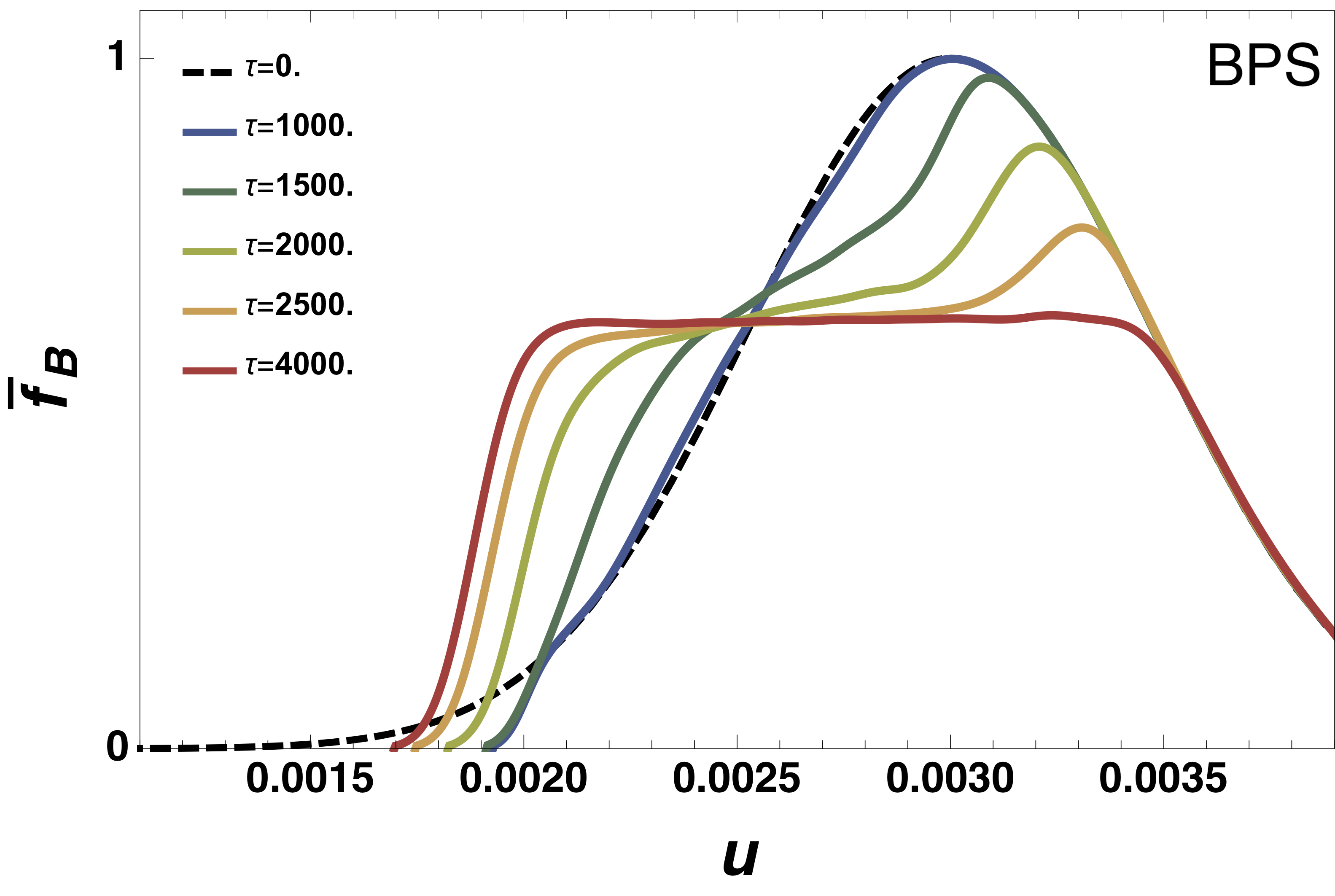}
\includegraphics[width=.32\textwidth,clip]{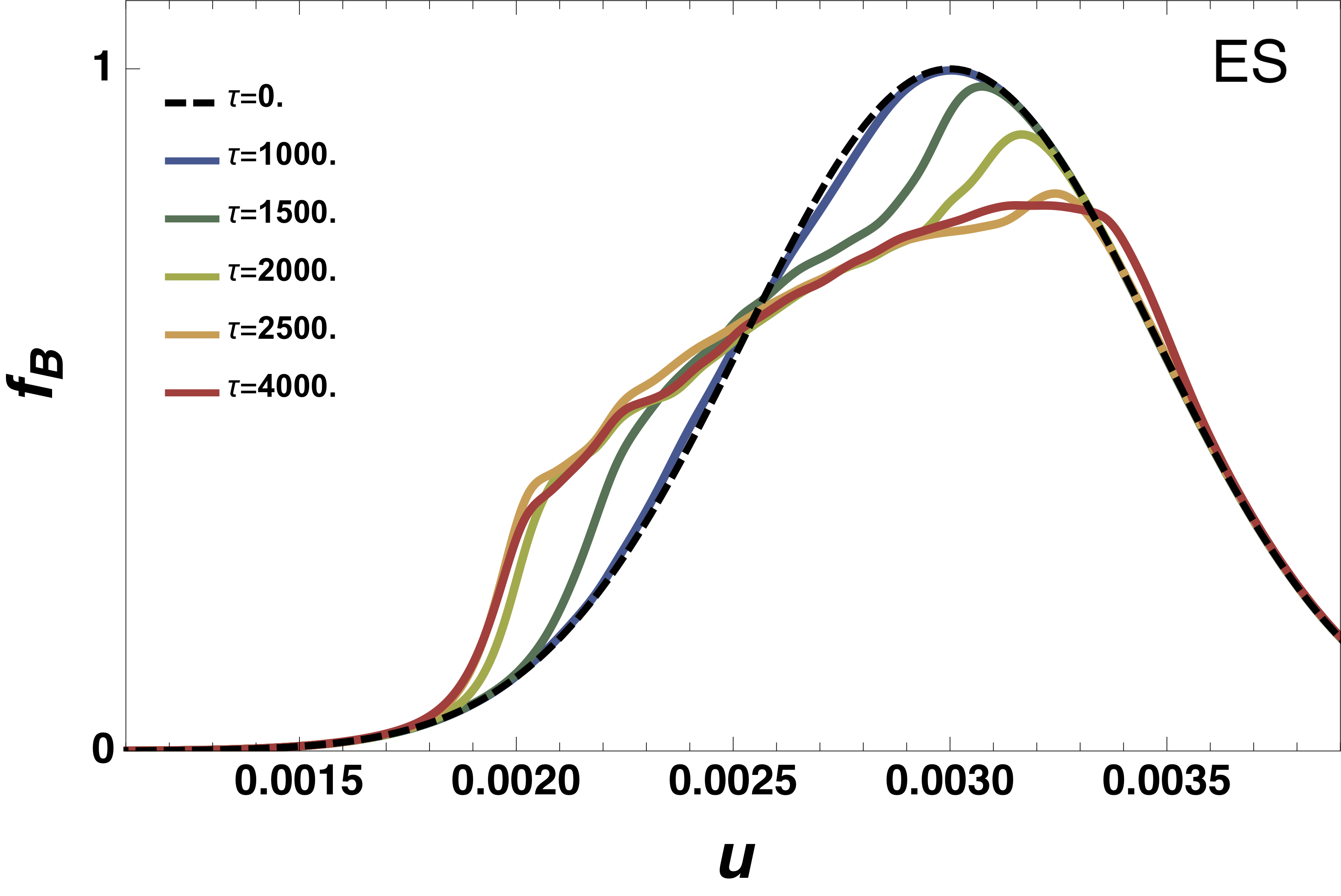}
\includegraphics[width=.32\textwidth,clip]{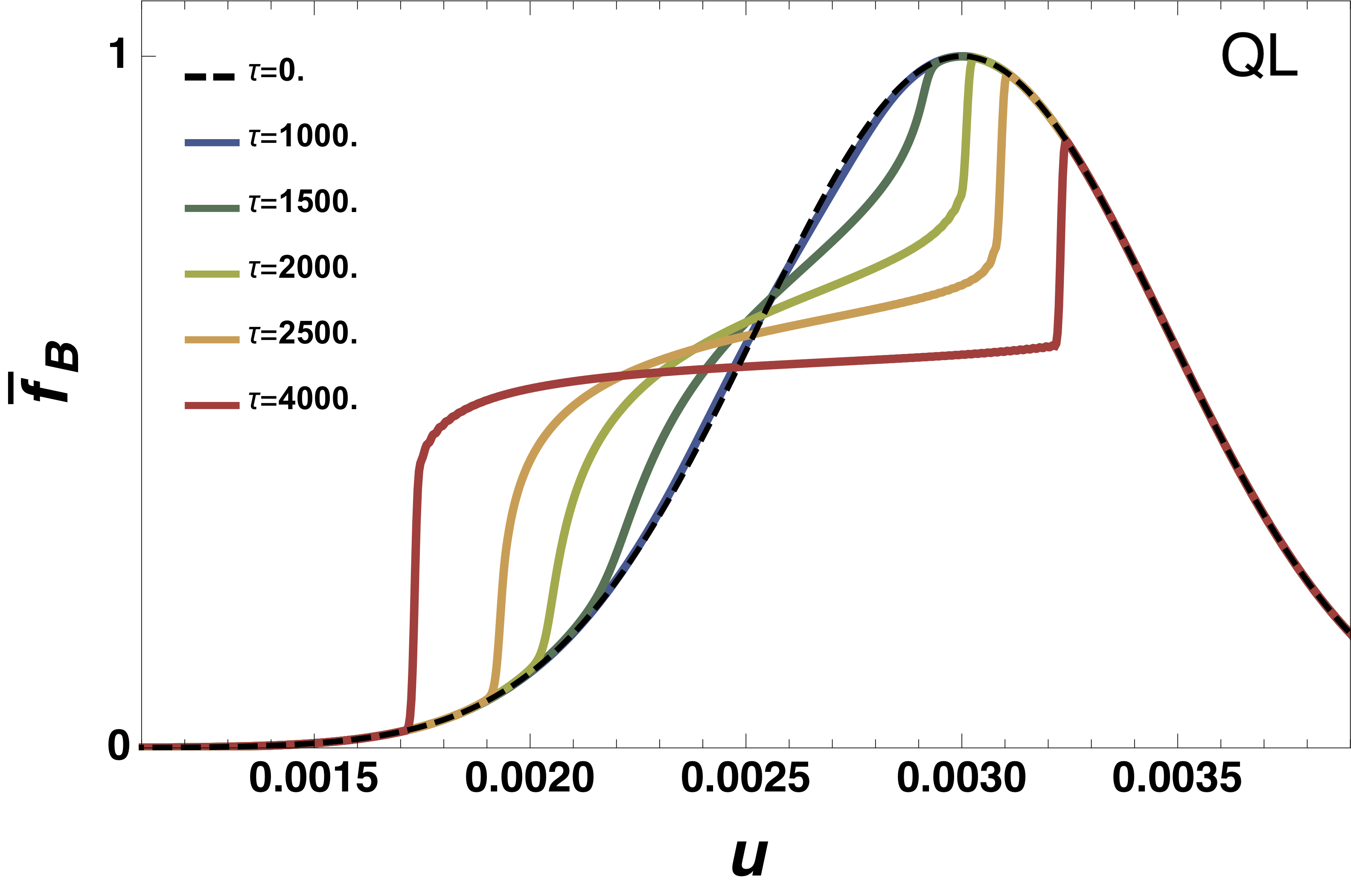}\vspace{-7mm}
\caption{\it \footnotesize DF evolution for: level 0, \eref{mainsys1} (left); level 1, \eref{deqexf} (center); level 2, \eref{QL_u} (right).
\label{figfb}}
\end{figure}

\vspace{-5mm}
\paragraph{Numerical results} We set a reference case of a Gaussian beam and $60$ modes corresponding to a scenario with Kubo number $\mathcal{K}=\tau_{ac}/\tau_{b}\sim0.02$. Simulations of \eref{mainsys1} are in \figref{figsetup} outlining the avalanche excitation of linear stable modes and the profile flattening (\figref{figfb}, left).
\vspace{-4mm}
\vspace{-3mm}
\vspace{-4mm}

The evolution of $\bar{f}_B$ for the various approximation levels is depicted in \figref{figfb}. The ES scheme (diagonal reduction) well reproduces the dynamics until saturation time scale, then mode-mode interaction ($k'\neq k$) becomes relevant and it loses predictivity. Regarding the QL evolution, we instead observe a retarded flattening formation, while the asymptotic plateau is well outlined.

\begin{figure}[ht!]
\centering
\includegraphics[width=0.32\linewidth]{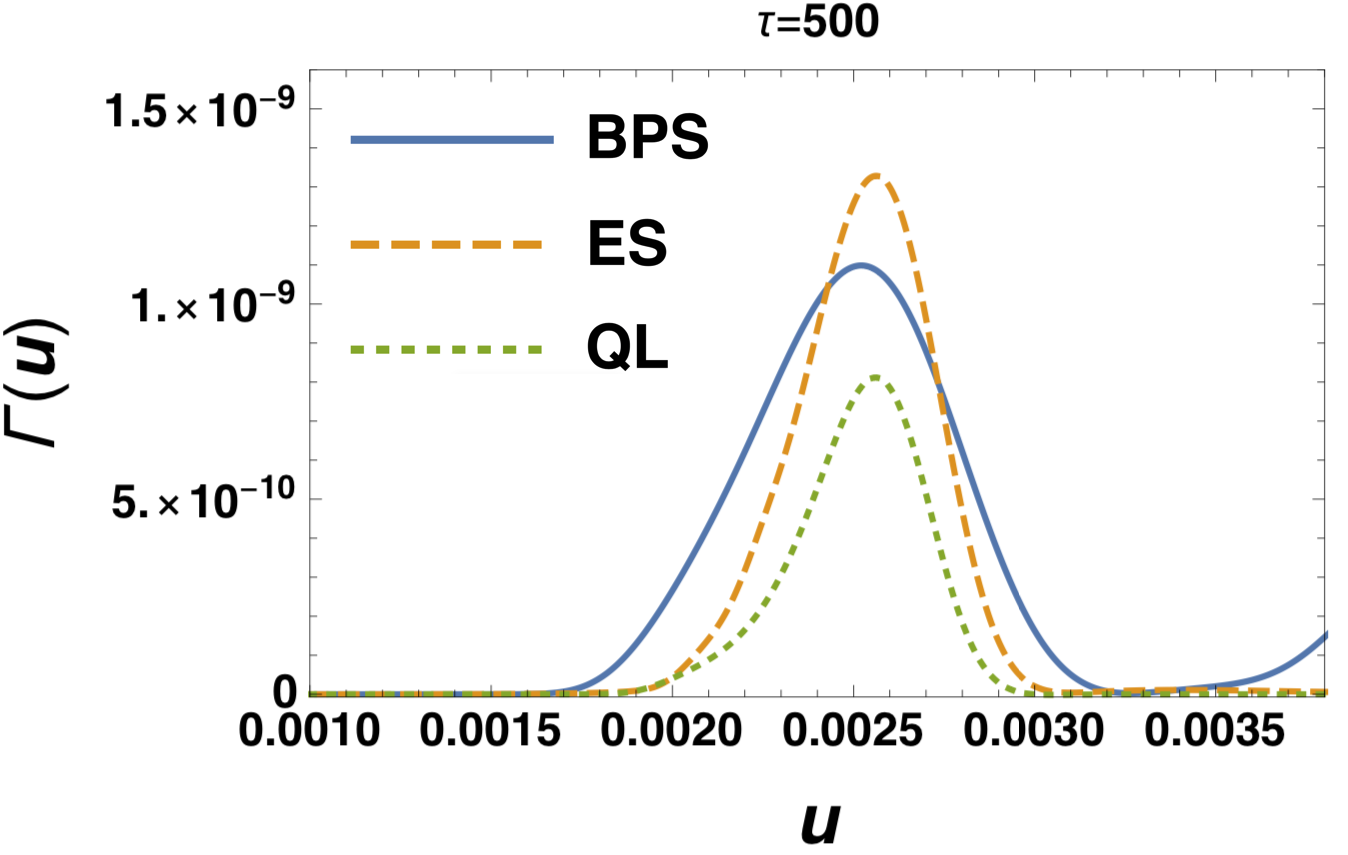}
\includegraphics[width=0.32\linewidth]{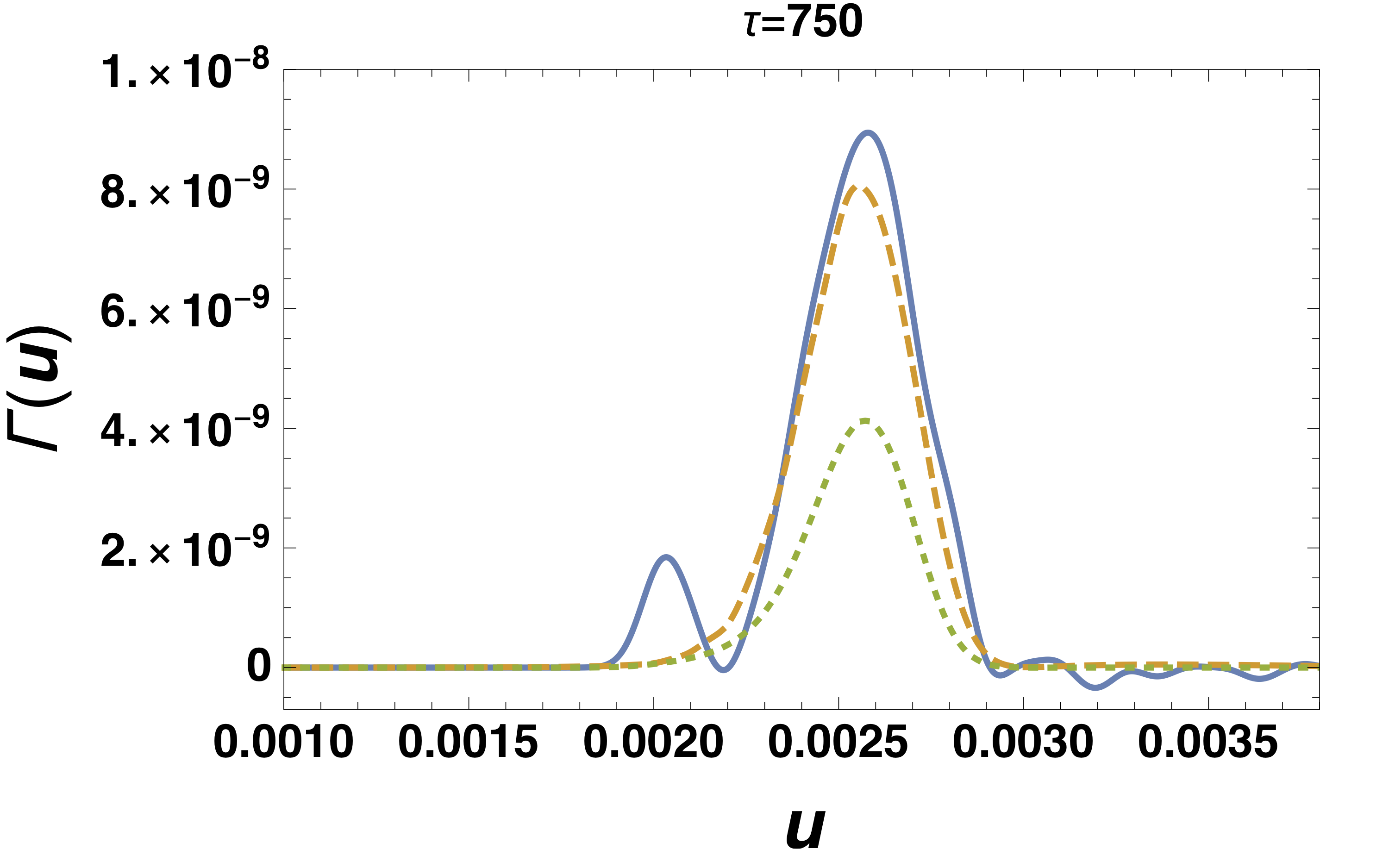}
\includegraphics[width=0.32\linewidth]{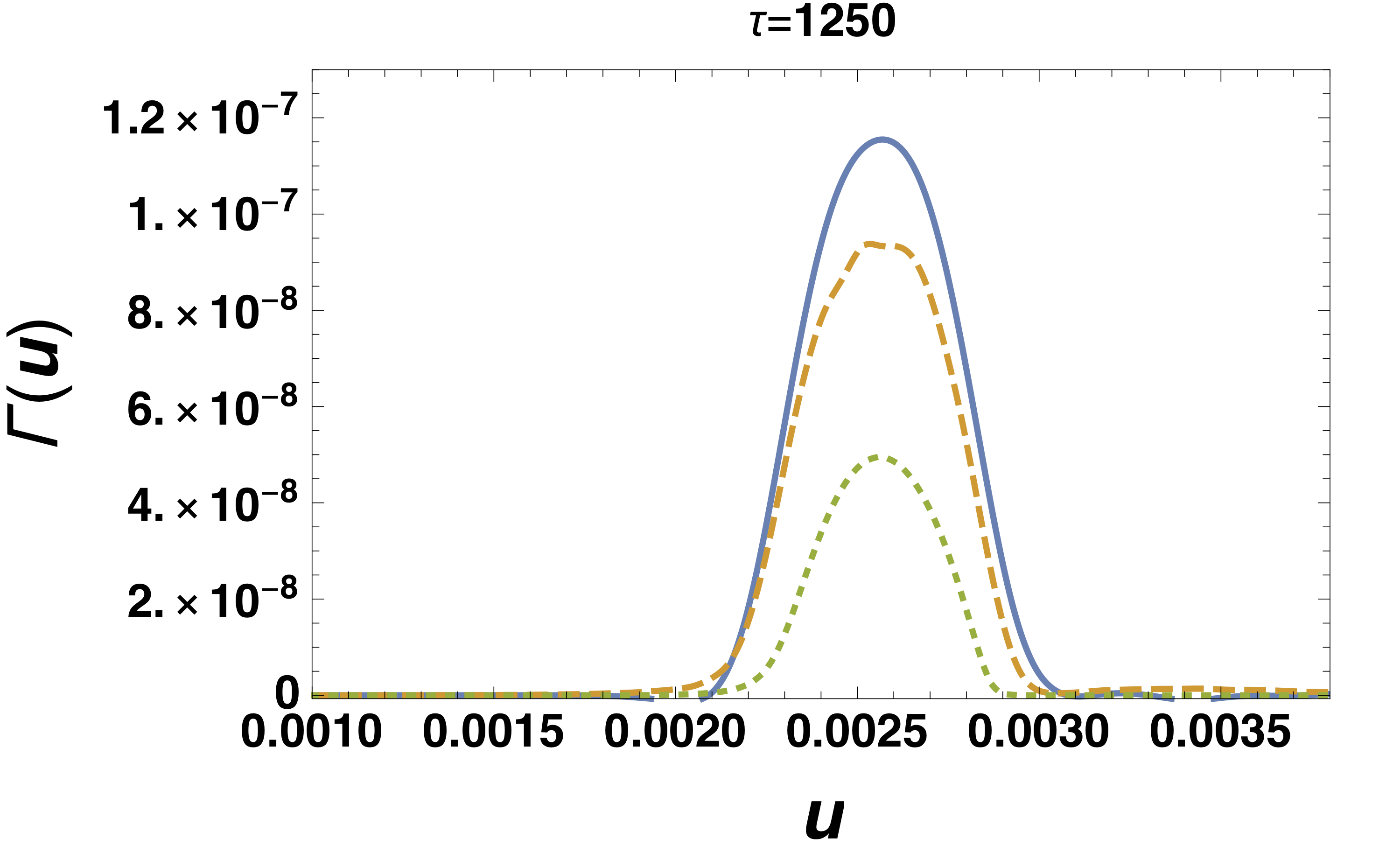}\\
\includegraphics[width=0.32\linewidth]{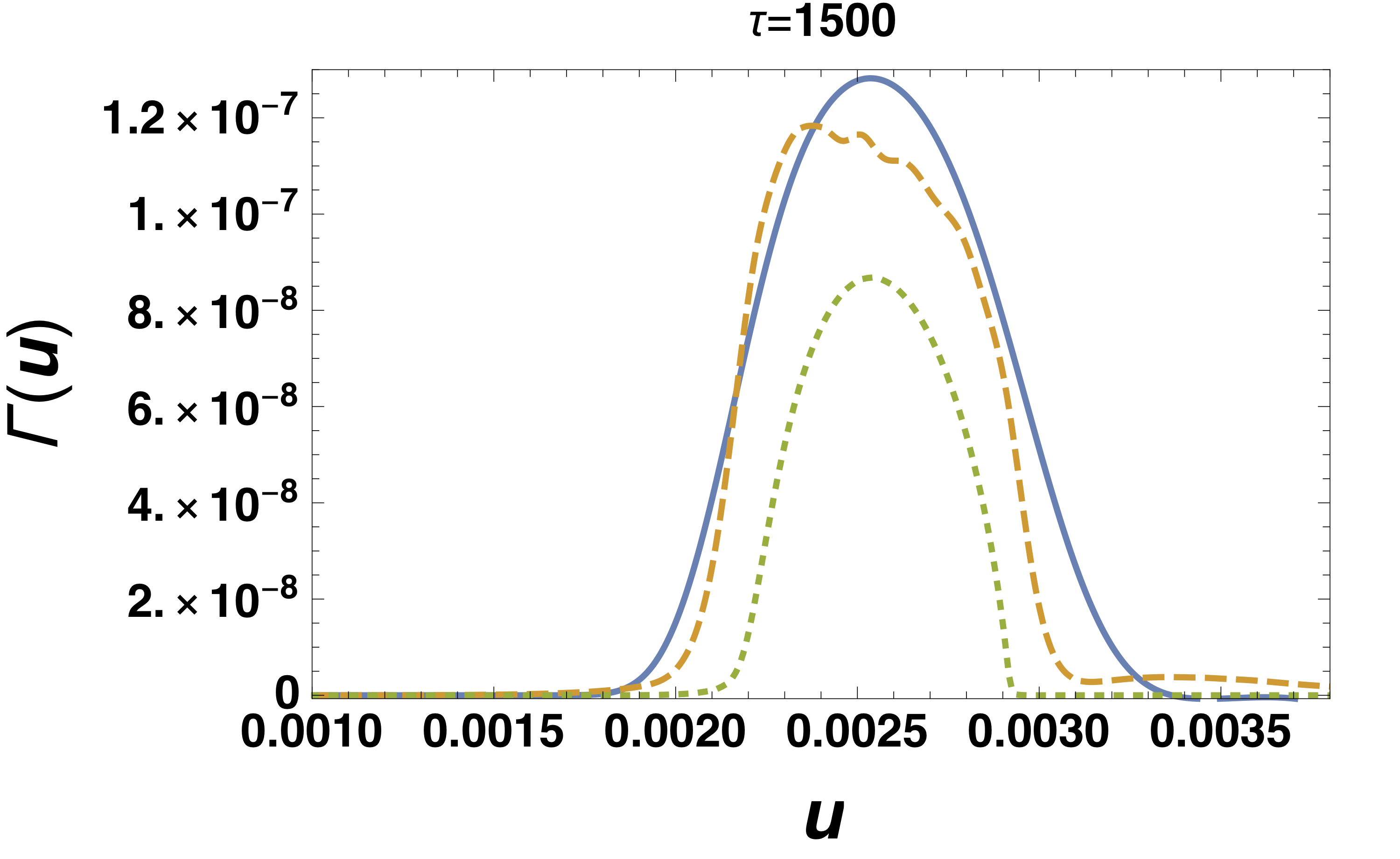}
\includegraphics[width=0.32\linewidth]{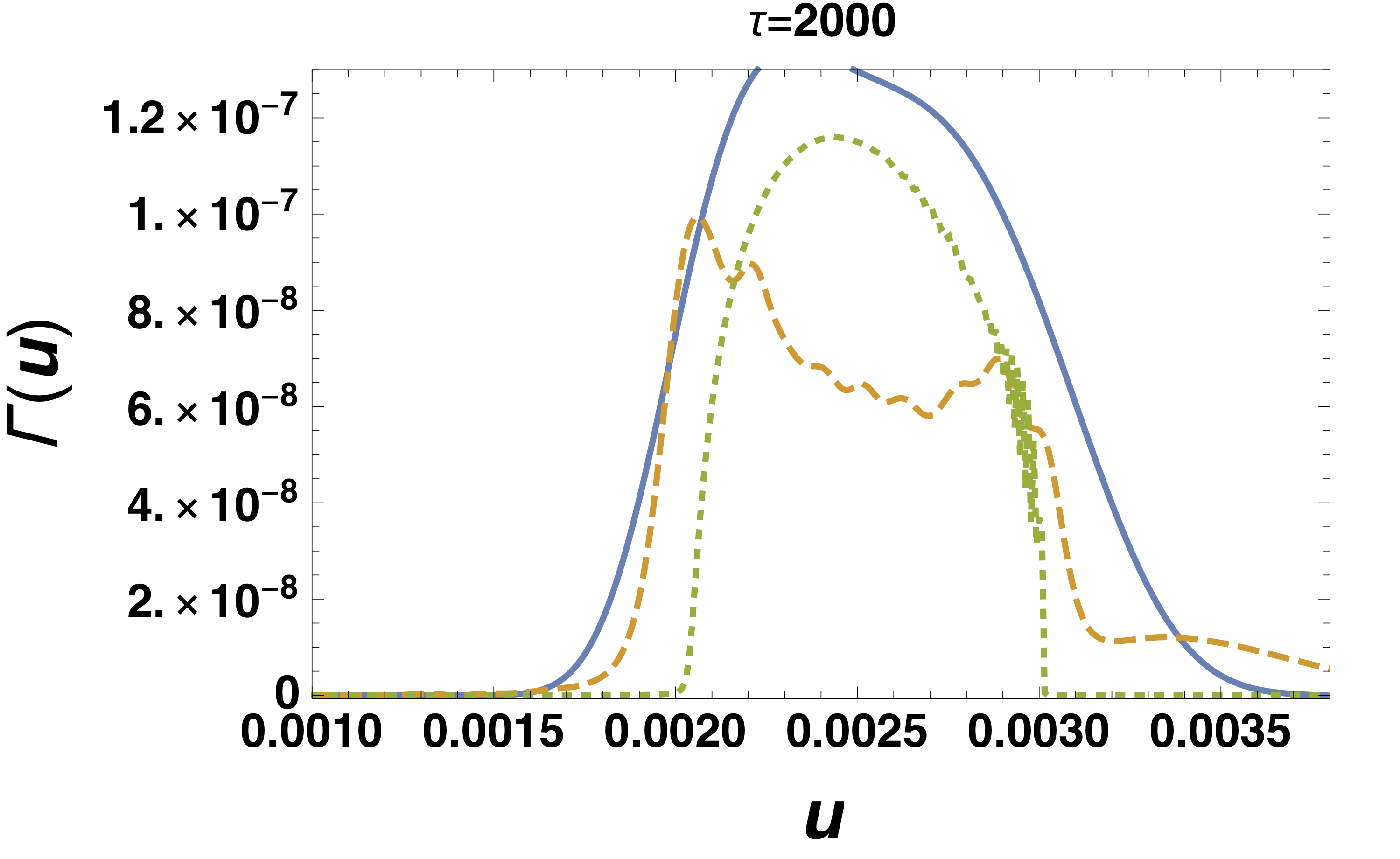}
\includegraphics[width=0.32\linewidth]{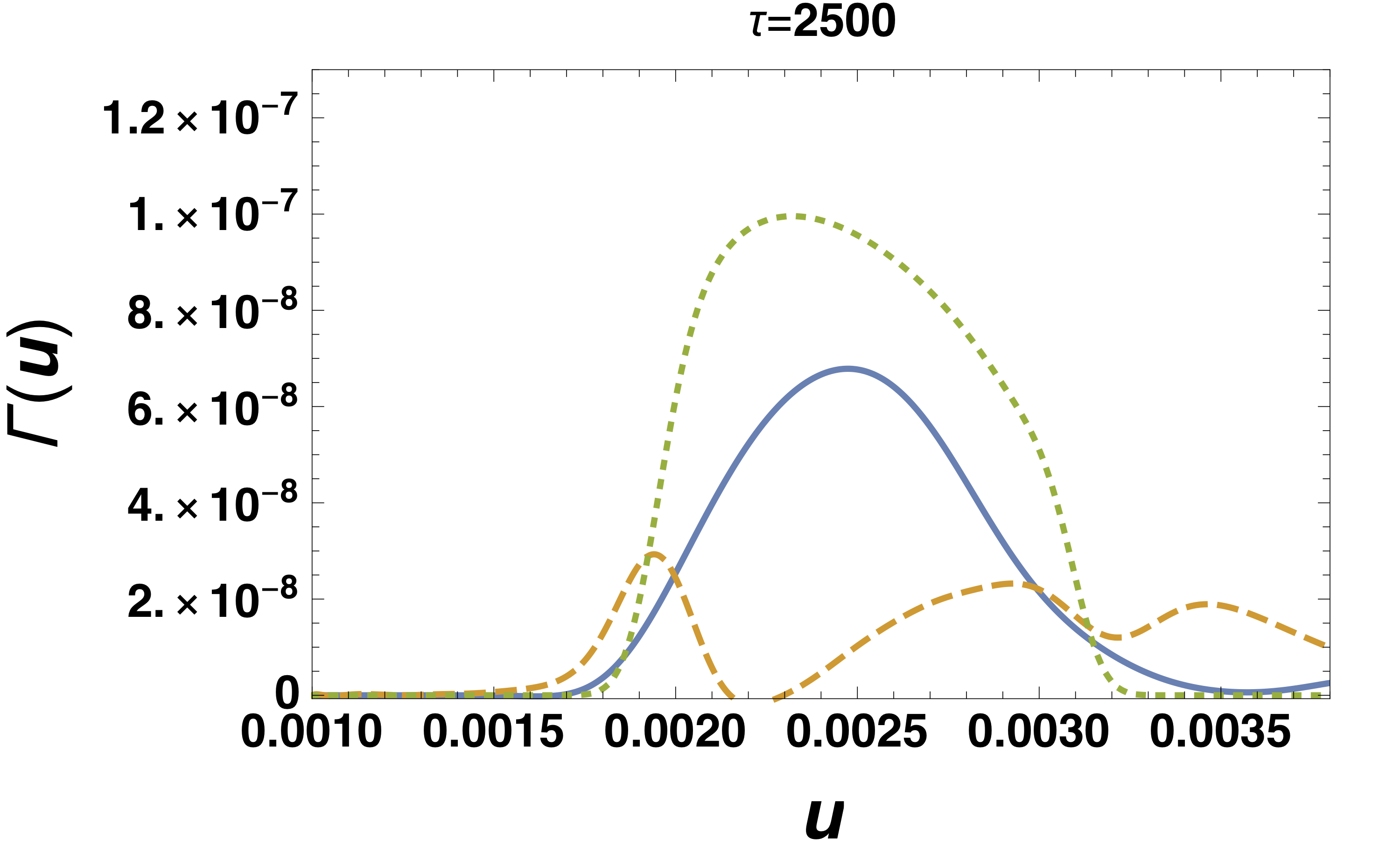}\vspace{-7mm}
\caption{\it \footnotesize Flux evolution ($\Gamma_{bps}$, $\Gamma_{es}$ and $\Gamma_{ql}$) for the different levels of approximation.
\label{figflux}}
\end{figure}

The fluxes $\Gamma_{bps}$ (level 0, \eref{dbnksdkjb}), $\Gamma_{es}$ (level 1, \eref{deqexf}) and $\Gamma_{ql}$ (level 2, \eref{QL_u}) are evaluated at different times (\figref{figflux}). Since fluxes correspond to the DF drive, the properties of $f_B$ discussed above are reflected in their evolution. In fact, for saturation time scales, the ES approximation well matches the self-consistent fluxes, while after saturation ($\tau\geq2000$) it loses predictivity and the QL model starts to be comparable to the N-body simulations.

\begin{figure}[ht!]
\centering
\includegraphics[width=0.315\linewidth]{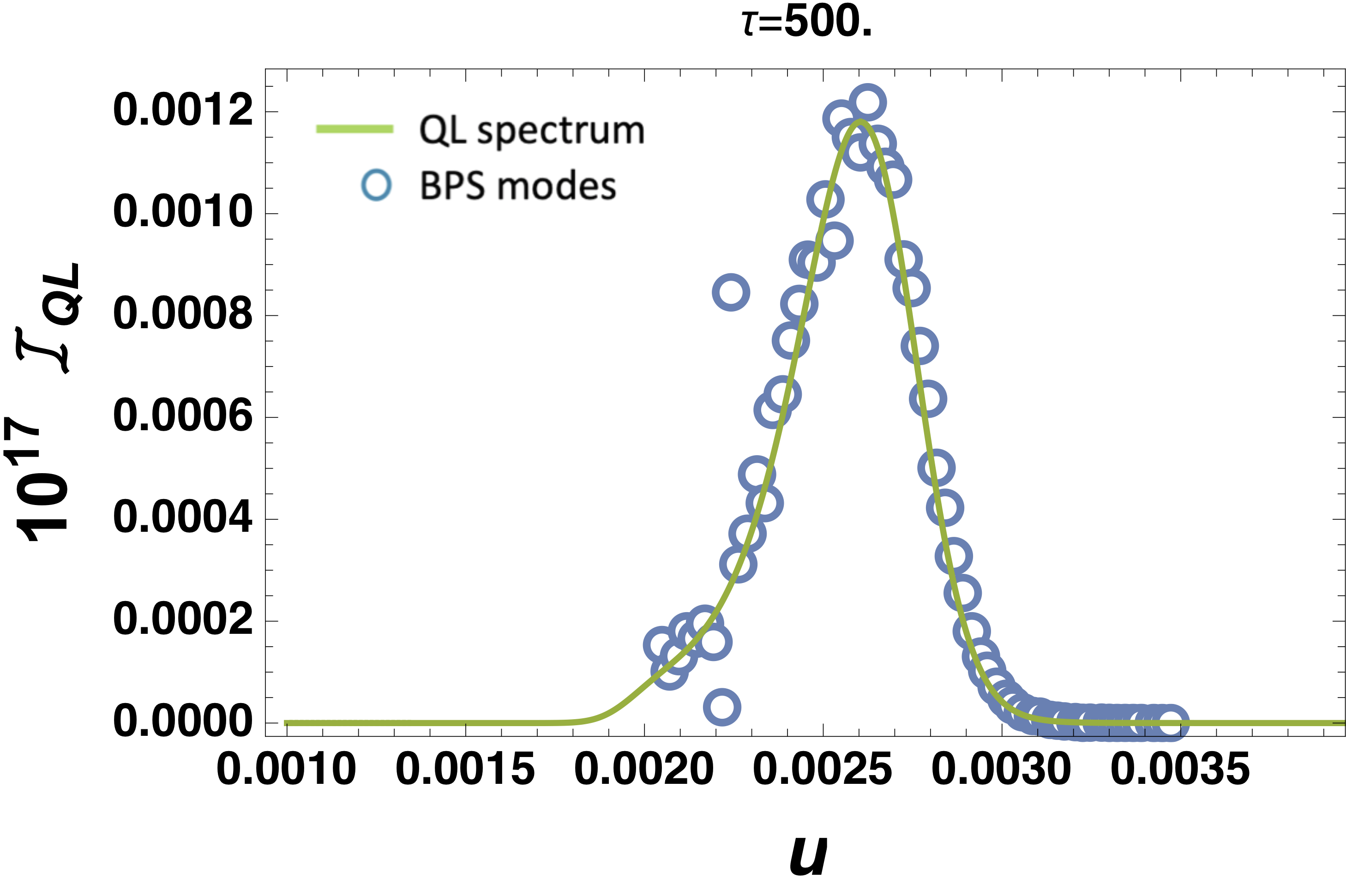}
\includegraphics[width=0.305\linewidth]{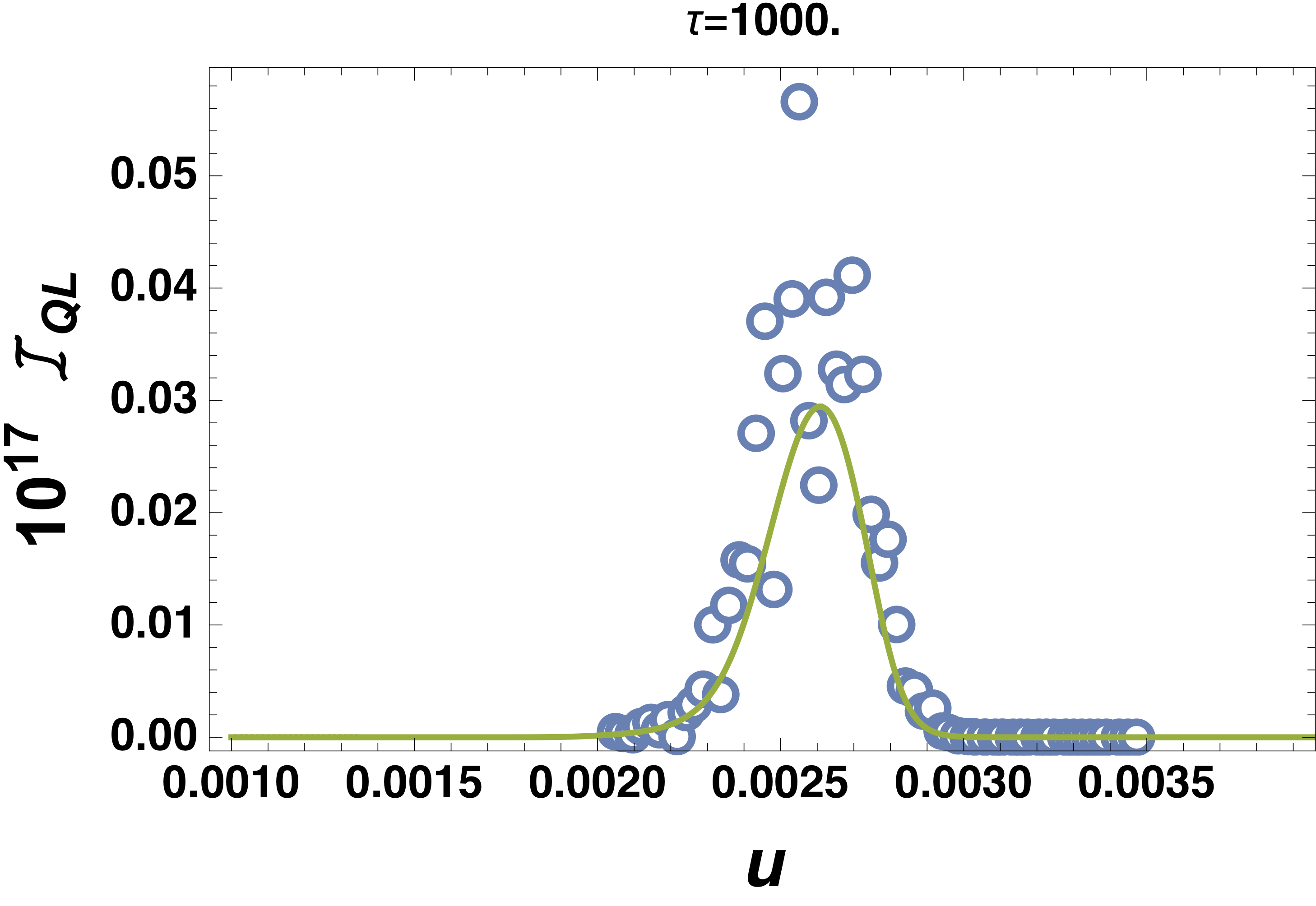}
\includegraphics[width=0.305\linewidth]{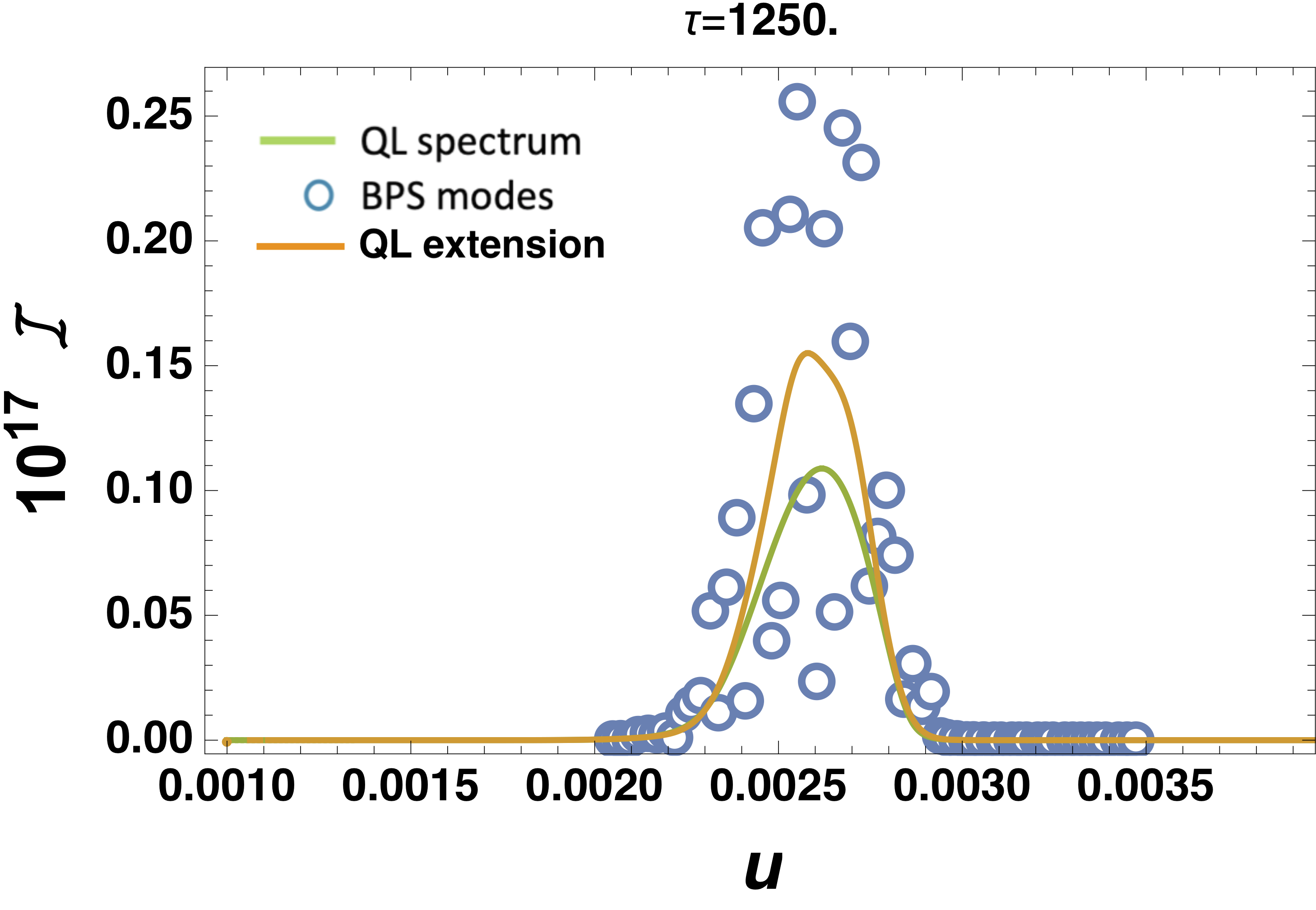}\\
\includegraphics[width=0.31\linewidth]{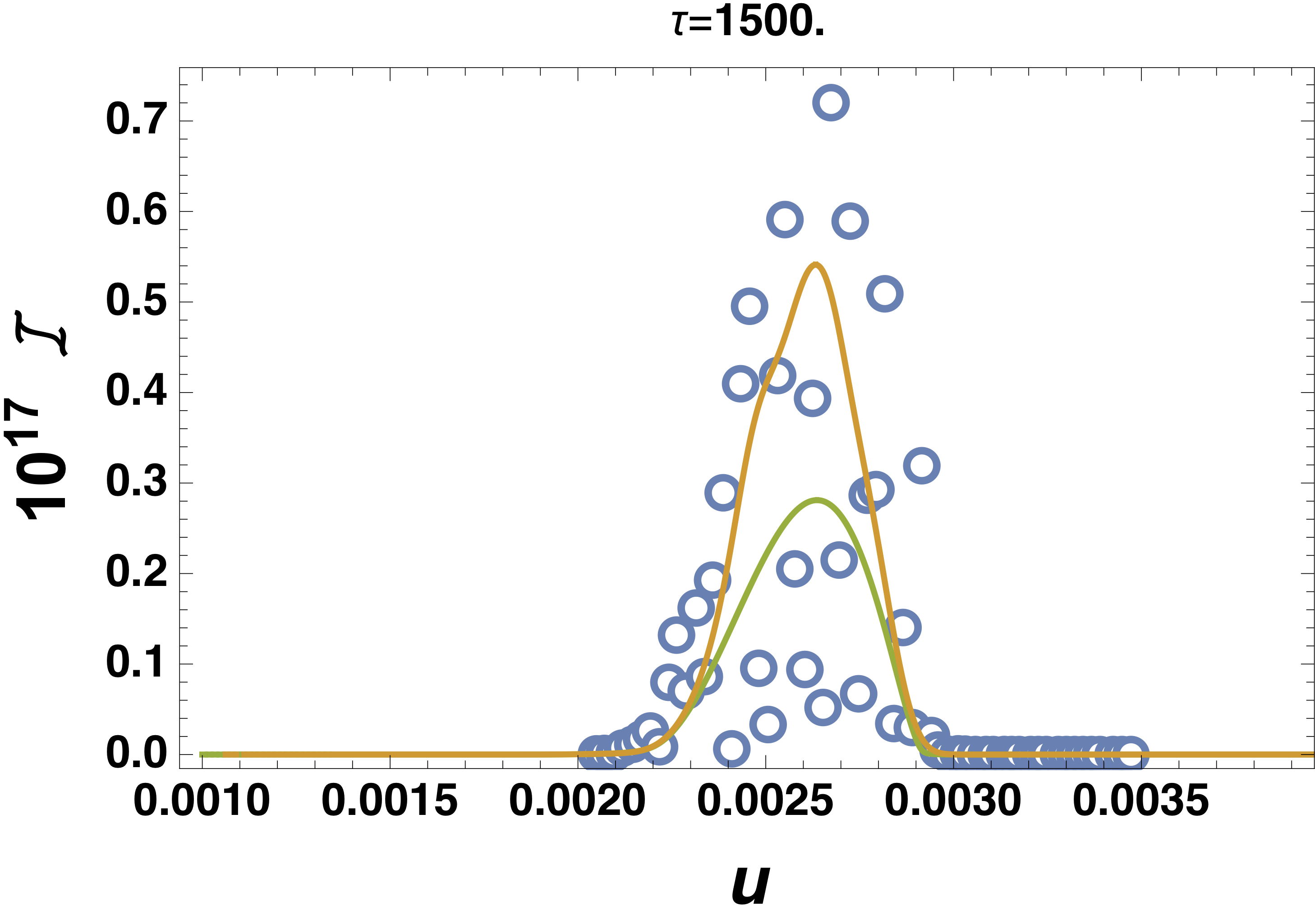}
\includegraphics[width=0.31\linewidth]{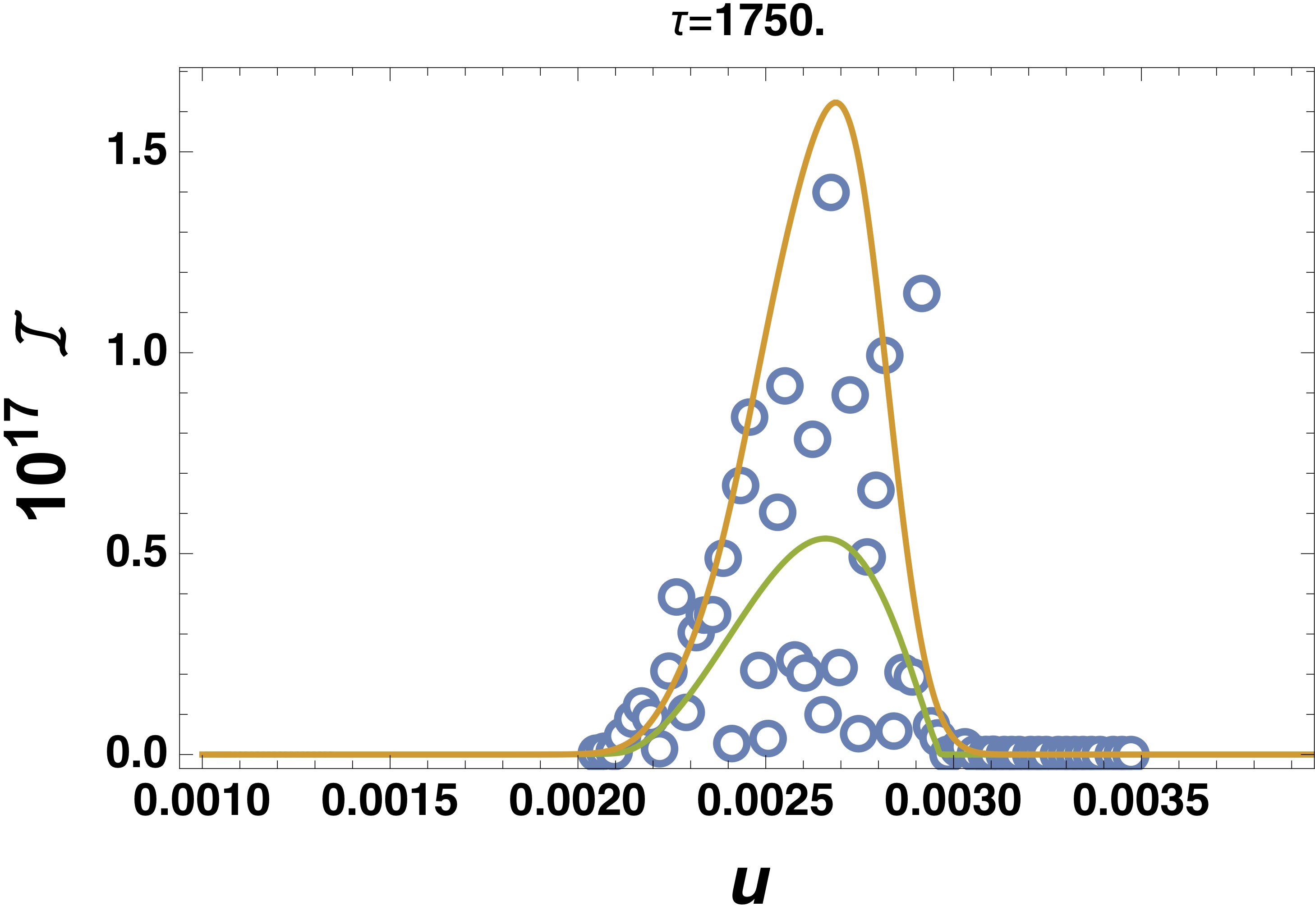}
\includegraphics[width=0.305\linewidth]{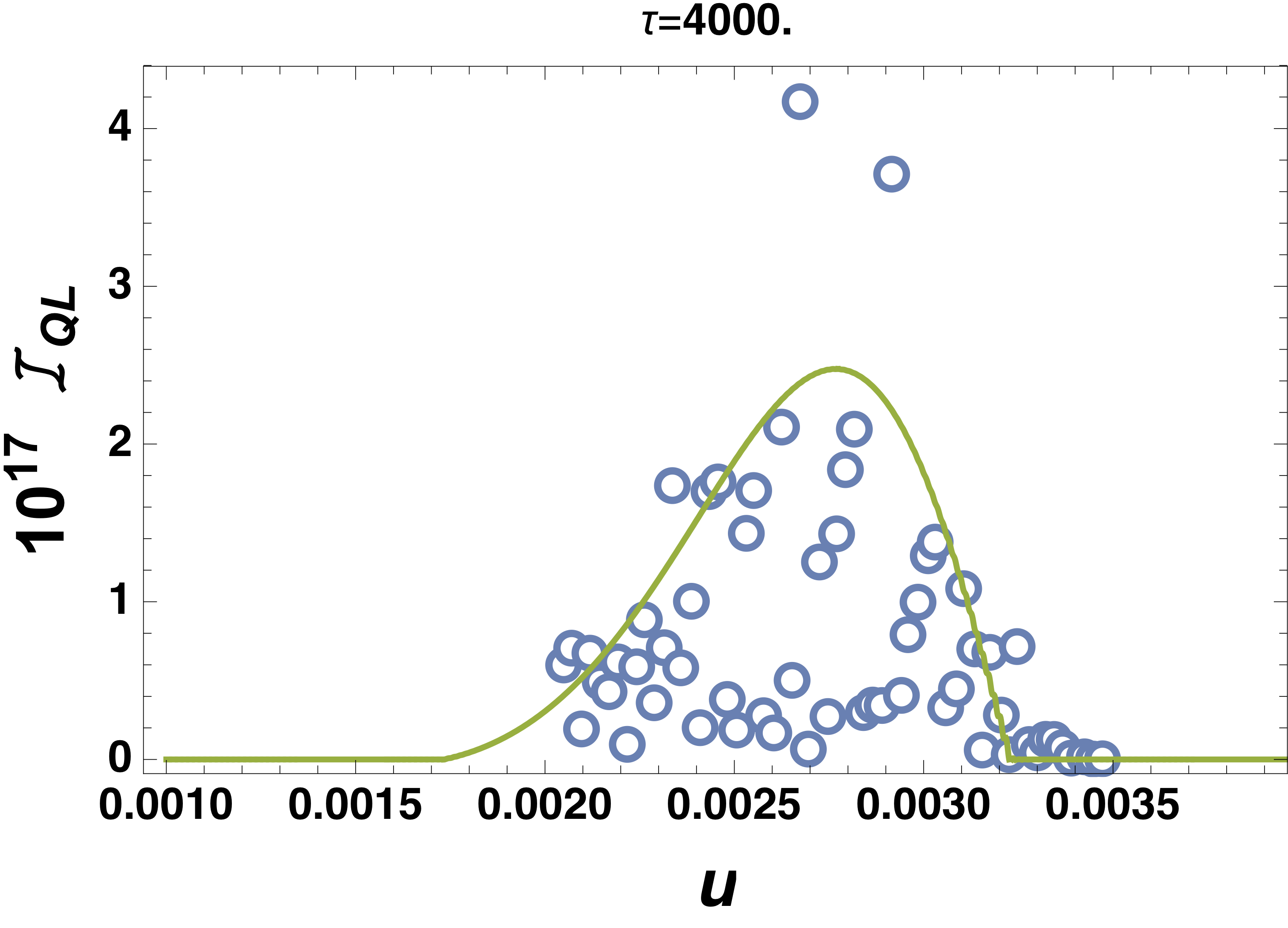}\vspace{-7mm}
\caption{\it \footnotesize QL spectral evolution from \eref{QL_u} (green) and  from \eref{snknlkn} (orange). Bullets are modes from \eref{mainsys1}.
\label{figspaect}}
\end{figure}

We conclude by plotting the spectral evolution compared to the mode evolution of the self consistent simulations (\figref{figspaect}). The QL model (\eref{QL_u}) is not predictive for the temporal meso-scales due to the non-pure diffusive character of the transport, while it properly envelope the discrete spectrum for late stages. Moreover, plotting the first order QL spectral extension (\eref{snknlkn}) for the late linear phase, we outline how it properly enhances the instantaneous growth rate curing the mesoscale spectral evolution in the pre-saturation regime.
\vspace{-5mm}
\vspace{-8mm}

\paragraph{Concluding remarks}
Our analysis, based on $N$-body simulations as reference term for establishing the predictivity of different VP equation approximations, fixed a precise hierarchy related to different time scales. While the QL model is predictive in the late evolution, it fails in the temporal meso-scale, where the diagonal VP formulation appears as very reliable to account for the spectrum saturation. The latter is, thus, the most appropriate paradigm when the isomorphism between the BPS and the fast ions interacting with with Alv\'enic modes is implemented.

\tiny This work has been carried out within the framework of the EUROfusion Consortium [ER Project MET (CfP-AWP19-ENR-01-ENEA-05)] and has received funding from the Euratom research and training programme 2014-2018 and 2019-2020 under grant agreement No 633053. The views and opinions expressed herein do not necessarily reflect those of the European Commission.

\end{document}